\documentclass[twocolumn,
preprintnumbers,amsmath,amssymb,prd,superscriptaddress]{revtex4}

\usepackage{amsmath}
\usepackage{amsthm}
\usepackage{amssymb}
\usepackage{amscd}
\usepackage{amsfonts}
\usepackage{array}
\usepackage{graphicx}
\usepackage{tipa}

\def\R{{\mathbb{R}}}

\def\si{{\sigma}}

\def\mN{{\mathcal{N}}}

\def\si{{\sigma}}
\def\G{{\Gamma}}

\def\G{\Gamma}

\def\>{{\geq }}
\def\<{{\leq }}

\def\Iv{I^v_{\rm vect}}
\def\Ie{I^e_{\rm hyper}}

\def\Ze{Z^e_{\rm hyper}}

\newcommand{\bH}{\ensuremath{\mathbb{H}}}

\newcommand{\bR}{\ensuremath{\mathbb{R}}}

\newcommand{\bT}{\ensuremath{\mathbb{T}}}

\newcommand{\scC}{\ensuremath{\mathcal{C}}}

\newcommand{\scE}{\ensuremath{\mathcal{E}}}

\newcommand{\scG}{\ensuremath{\mathcal{G}}}

\newcommand{\scN}{\ensuremath{\mathcal{N}}}

\newcommand{\scS}{\ensuremath{\mathcal{S}}}

\newcommand{\scW}{\ensuremath{\mathcal{W}}}

\newcommand{\bea}{\begin{equation}\begin{aligned}}
\newcommand{\eea}{\end{aligned}\end{equation}}
\newcommand{\beq}{\begin{eqnarray}}
\newcommand{\eeq}{\end{eqnarray}}

\begin{document}

\preprint{PUPT-2408}

\title{Emergent 3-manifolds from 4d Superconformal Indices}


\author{Yuji Terashima}
\affiliation{Department of Mathematics, Tokyo Institute for 
Technology, Tokyo 152-8551, Japan}

\author{Masahito Yamazaki}
\affiliation{Princeton Center for Theoretical Science,
 Princeton University, Princeton NJ 08540, USA}

\date{\today}

\begin{abstract}

We show that the smooth geometry of a hyperbolic  3-manifold
emerges from a classical spin system defined on a 2d discrete lattice, 
and moreover show
that the process of this ``dimensional oxidation'' 
is equivalent with the dimensional reduction of a supersymmetric
gauge theory from 4d to 3d.
More concretely, we propose
an equality 
between (1) the 4d superconformal index of a 4d $\mathcal{N}=1$
superconformal quiver gauge theory described by a bipartite graph on
 $T^2$ and (2) the partition function of a classical
integrable spin chain on $T^2$. 
The 2d spin system is lifted to a hyperbolic 3-manifold after the
dimensional reduction and the Higgsing of the 4d gauge theory.

\end{abstract}

\maketitle


{\it Introduction.}---
The concept of spacetime has been of crucial importance 
in our understanding of Nature.
However, in the theory of quantum gravity,
it is widely believed that even the notion of classical
spacetime
is of secondary nature, and emerges from 
a more fundamental structure. One proposal for such a structure
is the spin network \cite{SpinNetwork}, 
a spin system defined on a discrete lattice.

In a different line of development,
more recently there have been important
developments in
supersymmetric gauge theories suggesting that
the spacetime geometry could be traded for 
another ``internal'' geometry.
This has been discussed for a 
class of supersymmetric gauge theories
compactified on a compact curved manifold,
which is thought of as the Euclidean version of the spacetime for 
the theory.
The idea is simple; we begin with a $D$-dimensional field theory
and compactify the theory on a class of $d_1$-dimensional manifolds
$\mathcal{C}$. The resulting $d_2$-dimensional theory is defined on a 
fixed $d_2$-dimensional compact manifold $\mathcal{S}$,
where $d_1+d_2=D$. 
We could instead first compactify on $\mathcal{S}$,
and then we have a $d_1$-dimensional theory on $\mathcal{C}$.
Thus we have a correspondence between 
the $d_2$-dimensional field theory on $\mathcal{S}$
and the $d_1$-dimensional field theory on $\mathcal{C}$.

While the idea itself is rather general,
in practice it is a rather difficult problem
to make a precise identification between the
observables of the two theories,
since a quantity on one side could
take a rather different form on the other.
A successful example of such a quantitative
identification is the relation 
between  the $S^4$ partition function of 4d $\mathcal{N}=2$
superconformal field theories (SCFT) arising from
a compactification of 6d $(2,0)$ theory on 
a Riemann surface $C$ \cite{Gaiotto:2009we} 
and a correlation function of
2d Liouville theory on $C$ \cite{Alday:2009aq}. 

The goal of this Letter is to unify
these two apparently unrelated ideas 
in supersymmetric gauge theories and gravity.
This gives new perspectives on the 
emergence of classical geometry,
and surprisingly the process has a counterpart in the
 supersymmetric gauge theory.

We  analyze the 4d superconformal index for quiver gauge theories 
dual to toric Calabi-Yau 3-folds, and find that the 
4d index is equivalent to the partition function
of an integrable
spin system in 2d. 
We then discuss
dimensional reduction from the 4d index to the
3d partition function
of the supersymmetric gauge theory;
this is to take a particular limit of the 4d index.
Surprisingly, on the 2d spin system side
this limit is translated into a limit where
classical/quantum geometry of a hyperbolic
3-manifold
\footnote{
The quantum geometry is captured by an $SL(2)$ Chern-Simons theory.
The role of the Planck constant is played by the inverse of the level 
$t$ of the Chern-Simons theory, and the classical geometry is reproduced
in the saddle point approximation of the limit $t\to \infty$.
Fluctuations around the saddle point gives a number of interesting
enumerative invariants for 3-manifolds, for example
the Reidemeister-Ray-Singer torsion.
}
emerges from the 2d lattice. 
In other words, emergent geometry, which is nothing but a dimensional
oxidation \footnote{
The number of dimensions decreases in dimensional reduction.
Dimensional oxidation is the opposite, where
the number of dimensions increases in the process.
}
in our context, is translated into a dimensional reduction in
the other description! What is novel about this story is that the
emergent geometry on one side, where fluctuations of the geometry
are described by quantum hyperbolic geometry, is translated into a simple
dimensional reduction on the other side, with a fixed background (compactification 
manifold); the fluctuation of the background is traded for
the fluctuation of the gauge theory degrees of freedom on a fixed background.

We expect that this is a general feature of the correspondence between
$d_1$ and $d_2$-dimensional theories mentioned above. Since we are
splitting the $D$-dimensions into two, 
when we dimensionally {\it reduce} the theory on the $d_2$-dimensional side,
the dimension {\it increases} on the $d_1$-dimensional side
The theories in this Letter provide a concrete example of this phenomenon.


Further details will be presented in a separate publication \cite{YFullPaper}.

\begin{figure}[htbp]
\centering{\includegraphics[scale=0.4]{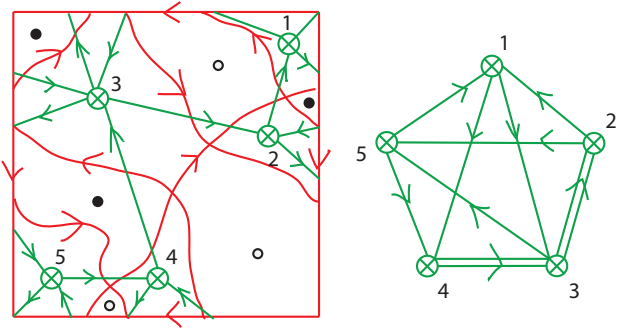}}
\caption{On a two-dimensional torus $\bT^2$ (the square
shown on the left figure)
we have a set of zig-zag paths, which divide the torus into
black/white regions represented by black/white vertices and 
uncolored regions represented by green vertices.
The right figure shows the corresponding quiver diagram.}
\label{fig3}
\end{figure}

{\it 4d versus 2d.}---
We begin with a 4d $\mathcal{N}=1$ quiver superconformal field theory obtained by probing toric Calabi-Yau
3-fold by $N$ D3-branes. In the following we take $N=2$.
Here a quiver gauge theory is a gauge theory defined from an
oriented graph (quiver); a vertex represents the $SU(2)$ gauge group and
an edge represents a bifundamental matter field.
In our case, the quiver is described by a set of zig-zag paths on $T^2$
(see Figure \ref{fig3})
\cite{Hanany:2005ss}. The paths represent the primitive normals of the toric diagram and divide $T^2$ into the 
polygonal regions, each of which is colored black/white if all the paths
around the region have counterclockwise/clockwise orientation and is uncolored
otherwise.
This determines the quiver diagram $\mathcal{G}$ or its dual $\mathcal{G}^*$, written on $T^2$. 
We denote the 
set of edges/faces/vertices of $\scG$ by $E, F, V$, $E/F$ is the
same as the 
colored/uncolored regions, and $\scG$ is a bipartite graph \cite{Hanany:2005ve}. 
In gauge theory language $V$ is the $SU(2)$ gauge group, $E$
bifundamental matter and $F$ the superpotential term. We denote the
endpoints of an edge $e\in E$ by $s(e), t(e)$.
In short, zig-zag paths determine the UV Lagrangian of our theory.

There is an ambiguity in the choice of zig-zag paths,
but assuming minimality all possible choices are related by the two moves \cite{Goncharov:2011hp},
representing Seiberg duality and integrating out massive bifudamental matters. These preserve the IR fixed point. 

Given a 4d $\scN=1$ SCFT we can 
define the superconformal index \cite{Romelsberger:2005eg} (defined on $S^3\times S^1$) by
\begin{align}
I(p,q)=\textrm{Tr}
\left[
(-1)^F p^{\frac{\scE+j_2}{3}+j_1}
 q^{\frac{\scE+j_2}{3}-j_1}
\right] ,
\end{align}
where the index is taken over the Hilbert space on $S^3$,
and  $F, \scE, j_1, j_2$ 
are the fermion number, the energy, and the spins under
$SO(4)=SU(2)\times SU(2)$ 
rotation symmetry \footnote{We can also include flavor chemical
potentials, and this will shift the R-charge when dimensionally reduced
to 3d.}.
This is essentially the famous Witten index, except that we here include
the chemical potentials for all the possible operators commuting with
the  supercharge.
Since the index is independent of the parameters of the Lagrangian \cite{Romelsberger:2005eg},
we can compute the index in the free field limit and
we have
\bea
I= \int _{|z_v|=1} \prod_{v \in V} \frac{dz_v}{z_v}
\prod_{v\in V} \Iv (z) 
\prod_{e \in E} \Ie (z;R) ,
\label{Iintegral}
\eea
where the integral is over the Cartan $(z_v, z_v^{-1}) \in U(1)_v\subset
SU(2)_v$ of the
gauge group
at vertex $v$.
The contribution from a gauge group at $v\in V$ is
\bea
\Iv (z)=\frac{\kappa}{2} 
\prod_{\epsilon=\pm 1}\G (z_{v}^{2\epsilon} ;p,q)^{-1},
\eea
and that from a bifundamental at $e\in E$ is
\bea
\Ie (z;R)=\prod_{\epsilon_1, \epsilon_2=\pm 1}\G ((pq)^{\frac{R_e}{2}}
z_{s(e)}^{\epsilon_1} z_{t(e)}^{\epsilon_2};p,q) ,
\eea
where $R_e$ is the R-charge for the bifundamental, satisfying 
the conditions that the superpotential has R-charge 2 and
that the $\beta$-functions vanish \cite{Hanany:2005ss}:
\begin{align}
\sum_{e: \textrm{ around } f} R_e=2, \quad
\sum_{e: \textrm{ incident to } v} (1-R_e)=2,
\label{R-charge}
\end{align}
for all $f\in F, v\in V$.
In these expressions we used an elliptic gamma function
\beq
\Gamma(x;p,q)=\prod_{j,k\ge 0} \frac{1-x^{-1} p^j q^k}{1-x p^j q^k},
\label{Gammadef}
\eeq
and $\kappa:=\prod_{j\ge 0}(1-p^{j+1})(1-q^{j+1})$.

Our key observation is that the index of our 4d theory is identified with 
the partition function of the Bazhanov-Sergeev spin model 
\cite{Bazhanov:2010kz} (see also  \cite{Bazhanov:2011mz})
under the following identification: \footnote{
There are in fact differences between the two. 
First, our model is defined on $T^2$, on the other hand  
their model is defined on $\R^2$.
Second, we do not include a spin-independent
normalization factor $\kappa (\alpha)$ 
in the weight of Bazhanov-Sergeev spin model. 
This only changes the overall normalization of the partition function
({\it i.e.} only the normalization outside the integral of
\eqref{Iintegral}), and our index is still invariant under the double
Yang-Baxter move \cite{YFullPaper}.
}
$$
\begin{array}{|>{$}c<{$}|>{$}c<{$}|}
\hline 
4d gauge theory & spin model \\ \hline\hline
quiver diagram & spin lattice \\ \hline
Cartan variable & spin variable \\ \hline
1-loop determinant & Boltzmann weight \\ \hline 
R-charge & spectral parameter \\ \hline
4d index & partition function \\ \hline
\end{array}
$$
In this correspondence, 
the invariance of the 4d index under Seiberg duality 
\cite{Dolan:2008qi} 
results from the star-triangle relation in 
the Bazhanov-Sergeev spin model \cite{Spiridonov:2010em}.
This spin system can be reduced to many known integrable models, for
example
the chiral Potts model \cite{Bazhanov:2010kz}.
It is known that the partition function of the chiral Potts model gives
a special value of Jones polynomial and HOMFLY (Hoste-Oceanu-Millett-Freyd-Lickorish-Yetter)
polynomial for a link
whose projection gives the zig-zag paths. This suggests that the
reduction process, which is different from the dimensional reduction
discussed in this Letter, connects the 4d quiver gauge theory to the 3d
Chern-Simons theory on the link complement. It would be interesting to give a
 gauge theory interpretation of this novel reduction.

Our 4d gauge theory has a brane realization in terms of $N$ D5-branes
and an NS5-brane \cite{Imamura}; in the notation of introduction,
we have $D=6, d_1=2, d_2=4, \scC=T^2$ and $\scS=S^3\times S^1$
\footnote{The topology of
$\scC$ is unique, however there is an extra ingredient, an NS5-brane,
which wraps a general Riemann surface.}.
However, it should be kept in mind that our correspondence 
has important differences from \cite{Alday:2009aq}
and their variants. For example, our 2d spin system is a {\it classical}
spin system, whereas in \cite{Alday:2009aq} the 2d system is the quantum
Liouville theory.

{\it 3d versus 3d.}---
We now consider a dimensional reduction of our theory;
we take the radius of thermal $S^1$ to zero, and all the KK modes 
decouple.
In this limit, all the chemical potentials go to $1$, but 
we can take the limit while keeping their ratio finite.
Our limit is $\beta \to 0$ with \cite{Dolan:2011rp}
\beq
p=e^{-\beta (1+\eta)}, \quad q=e^{- \beta (1-\eta)},
\eeq
after which 4d $\scN=1$ theory on $S^3 \times S^1$ 
reduces to 3d $\scN=2$ theory on an ellipsoid $S^3_b$ \cite{Hama:2011ea}, where $b^2=\frac{1+\eta}{1-\eta}$.
In addition we Higgs the theory to the Cartan, by giving a VEV (Vacuum Expectation Value) 
to the vectormultiplet scalar of the
diagonal gauge group $U(1)_{\rm diag}\subset SU(2)^{|V|}$
and sending it to infinity.
The 4d index now reduces to
\bea
Z_{\rm 3d}(R)=\int \prod_{v\in V} d\sigma_v  \prod_{e\in E} Z^e_{\rm hyper}(\sigma;R),
\label{Z3d}
\eea
where 
\bea
\Ze (\sigma;R)=\frac{s_b\left(\sigma_{s(e)}-\sigma_{t(e)}+\frac{iQ}{2}(1-R_e)\right)}{s_b\left(\sigma_{s(e)}-\sigma_{t(e)}-\frac{iQ}{2}(1-R_e)\right)},
\eea
the variable $\sigma_v$ is the vectormultiplet scalar, and $s_b(x)$ is
the quantum dilogarithm function. The integral in \eqref{Z3d}
is taken over the real axis.
The deformation parameter $b$ plays the role of the quantum parameter of
the theory.
The result \eqref{Z3d} coincides with the partition 
function of the Faddeev-Volkov model 
\cite{Bazhanov:2007mh, VolkovAbelian},
which describes a discrete Virasoro symmetry. 

\begin{figure}[htbp]
\centering{\includegraphics[scale=0.4]{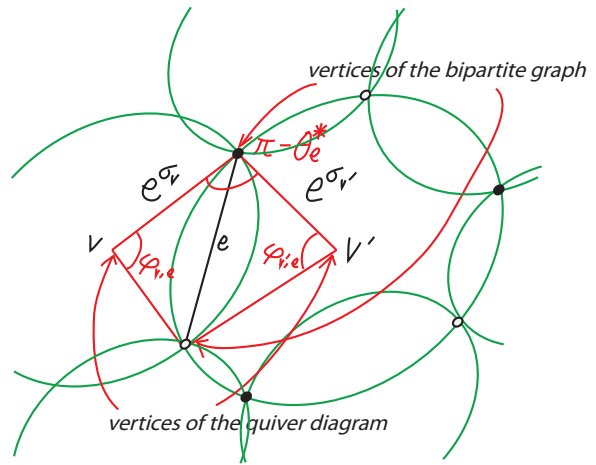}}
\caption{The projection of a non-ideal hyperbolic tetrahedron gives the triangle in the Figure.
 Our 3-manifold $M_R$ is obtained by gluing these tetrahedra.}
\label{fig1}
\end{figure}

In the semiclassical limit $b \to 0$, $S^3_b$ reduces to $\bR^2\times
S^1_b$ with $S^1_b$ of small radius $b$, and the theory effectively reduces to a 2d theory with all the
Kaluza-Klein modes included.
Indeed, (after rescaling $\sigma$) 
the 3d partition function reduces to an 
integral of the effective twisted superpotential 
$\scW_{\rm 2d}(\sigma;R)$:
\begin{equation}
Z_{\rm 3d}(R)=\int \prod_{v\in V} d\sigma_v \, \exp\left[\frac{1}{2\pi b^2} \scW_{\rm
 2d}(\sigma)\right],
\end{equation}
where
\begin{equation}
\begin{split}
\scW_{\rm 2d}(\si ;R)& =\sum_{e\in E} 
\left[l(\sigma_{s(e)}-\sigma_{t(e)}+i\theta^*_e)  \right.\\
& \quad \quad \qquad
\left. -l(\sigma_{s(e)}-\sigma_{t(e)}-i\theta^*_e)
\right].
\end{split}
\end{equation}
Here we defined $\theta^*_e=\pi (1-R_e)$ and $l(z)$ is defined 
from the classical dilogarithm function $\textrm{Li}_2(z)$ 
to be
\beq
l(z)=\textrm{Li}_2(-e^z)+\frac{1}{4}z^2 .
\eeq
An important result in  
\cite{Bazhanov:2007mh, Bobenko} states that the saddle point equation of 
the effective twisted superpotential 
can be interpreted as a gluing condition at 
the vertices of the quiver diagram 
of (non-ideal) hyperbolic tetrahedra whose 
projection to $\partial \bH^3$ is a triangle in 
Figure \ref{fig1}. 

The necessary and sufficient conditions for the existence of the
solution of the gluing condition 
is stated in \cite[Theorem 3]{Bobenko}.
The first condition is (in our notation)
 \begin{equation}
\sum_{f\in F} 2\pi =\sum_{e\in E} 2 (\pi-\theta^*_e).
\end{equation}
The second condition is that for a nonempty subset $F'$ of $F$ with $F\ne F'$ and the set $E'$ of all
      edges incident with any face of $F'$, we have 
\begin{equation}
\sum_{f\in F'} 2\pi < \sum_{e\in E'} 2 (\pi -\theta^*_e).
\end{equation}
These conditions follow from the 
conditions on the R-charge \eqref{R-charge};
the first (second) condition follows from
the sum of the first equation of \eqref{R-charge} over 
$f\in F$ ($f\in F'$). Note that each edge is adjacent to two faces.

After gluing tetrahedra we therefore have a 
3d hyperbolic manifold $M_R$ 
whose projection to $\partial \bH^3$ is 
combinatorially given by the bipartite graph on our $T^2$.
When the circle radii are all of the same value, 
the Legendre transform 
of the volume of $M_R$ 
with respect to the angles $\theta_e^*$
is
related with the prepotential of the
topological string theory on the dual toric Calabi-Yau 3-fold \cite{YFullPaper}.

The correspondence between 2d twisted superpotential and the 3d
classical hyperbolic geometry motivates us to propose
a quantum version of the correspondence:
the 3d partition function $Z_{\rm 3d}$ with finite $b$
compute the partition function of the 3d $SL(2)$
Chern-Simons theory, where $b$ is related to the level
$t$ by $t\sim 1/b^2$. Since 3d gravity is closely related with 3d
$SL(2)$
Chern-Simons theory \footnote{There are important differences between
the two. However, the difference is irrelevant for the consideration of this
paper since we only consider saddle point expansions around geometric
flat connections.}, this means that our 2d spin system is 
a version of the spin network for 3d gravity \footnote{
Our spins take continuous values, and 
transforms as a continuous representation of $SL(2)$.
This is in contrast with many literature
on spin networks,
where the spins are the discrete spins under the compact group $SU(2)$, not
the non-compact $SL(2)$ group as it really expected from 3d gravity.
}, i.e. a spin system defined on a discrete lattice which reproduces
3d gravity in a limit.

Finally, the results of this Letter
are reminiscent of two existing results in the literature.
The first is the 4d/2d relation between the 4d superconformal index
for Gaiotto theories and the correlation function of
2d TQFT (Topological Quantum Field Theory) \cite{TQFT}. The second is the 3d/3d relation 
between 3d $\mN =2$ theories and 3d $SL(2)$ Chern-Simons 
theories \cite{3d3d}.
It would be interesting to elucidate 
the precise relation between these and 
the results of this Letter. 

\medskip
\noindent{\bf Acknowledgment}:~
This research is supported in part by
the Grants-in-Aid for Scientific Research, JSPS (Y.~T.)
and by PCTS (M.~Y.). Y.~T. thanks H.~Fuji, Y.~Imamura, K.~Ito, K.~Nagao, 
M.~Shibata, S.~Terashima for helpful conversations. M.~Y. thanks JINR, Cambridge University and
University of Munich for hospitality.



\begin{thebibliography}{10}

\bibitem{SpinNetwork} 
  R.~Penrose, ``Angular momentum; an approach to combinatorial space
time,'' in {\it Quantum Theory and Beyond}, T. Bastin (ed.), Cambridge, 1971.


\bibitem{Gaiotto:2009we}
 D.~Gaiotto,
 arXiv:0904.2715 [hep-th].

\bibitem{Alday:2009aq}
 L.~F.~Alday, D.~Gaiotto and Y.~Tachikawa,
 Lett.\ Math.\ Phys.\  {\bf 91}, 167 (2010)
 [arXiv:0906.3219 [hep-th]].

\bibitem{YFullPaper}
 M.~Yamazaki, 
 ``Quivers, YBE and 3-manifolds,'' 
 JHEP {\bf 1205}, 147 (2012) [arXiv:1203.5784[hep-th]].

\bibitem{Hanany:2005ss}
 A.~Hanany and D.~Vegh,
 JHEP {\bf 0710}, 029 (2007)
 [arXiv:hep-th/0511063].

\bibitem{Hanany:2005ve}
 A.~Hanany and K.~D.~Kennaway,
 arXiv:hep-th/0503149;
 S.~Franco, A.~Hanany, K.~D.~Kennaway, D.~Vegh and B.~Wecht,
 JHEP {\bf 0601}, 096 (2006)
 [arXiv:hep-th/0504110].

\bibitem{Goncharov:2011hp}
 A.~B.~Goncharov and R.~Kenyon,
 arXiv:1107.5588 [math.AG].

\bibitem{Romelsberger:2005eg}
 C.~Romelsberger,
 Nucl.\ Phys.\  B {\bf 747}, 329 (2006)
 [arXiv:hep-th/0510060];
 J.~Kinney, J.~M.~Maldacena, S.~Minwalla and S.~Raju,
 Commun.\ Math.\ Phys.\  {\bf 275}, 209 (2007)
 [arXiv:hep-th/0510251].

\bibitem{Bazhanov:2010kz}
 V.~V.~Bazhanov and S.~M.~Sergeev,
 arXiv:1006.0651 [math-ph].
 
 \bibitem{Bazhanov:2011mz} 
 V.~V.~Bazhanov and S.~M.~Sergeev,
 Nucl.\ Phys.\ B {\bf 856}, 475 (2012)
 [arXiv:1106.5874 [math-ph]].

\bibitem{Dolan:2008qi}
 F.~A.~Dolan and H.~Osborn,
 Nucl.\ Phys.\  B {\bf 818}, 137 (2009)
 [arXiv:0801.4947 [hep-th]]; 
   V.~P.~Spiridonov and G.~S.~Vartanov,
  Commun.\ Math.\ Phys.\  {\bf 304}, 797 (2011)
  [arXiv:0910.5944 [hep-th]],
  arXiv:1107.5788 [hep-th];
  A.~Gadde, L.~Rastelli, S.~S.~Razamat and W.~Yan,
  JHEP {\bf 1103}, 041 (2011)
  [arXiv:1011.5278 [hep-th]].


\bibitem{Spiridonov:2010em}
 V.~P.~Spiridonov,
 arXiv:1011.3798 [hep-th].

\bibitem{Imamura}
 Y.~Imamura,
 JHEP {\bf 0612}, 041 (2006)
 [hep-th/0609163];
  Y.~Imamura, H.~Isono, K.~Kimura and M.~Yamazaki,
  Prog.\ Theor.\ Phys.\  {\bf 117}, 923 (2007)
  [hep-th/0702049];
 M.~Yamazaki,
 Fortsch.\ Phys.\  {\bf 56}, 555 (2008)
 [arXiv:0803.4474 [hep-th]].

\bibitem{Dolan:2011rp}
 F.~A.~H.~Dolan, V.~P.~Spiridonov and G.~S.~Vartanov,
 Phys.\ Lett.\  B {\bf 704}, 234 (2011)
 [arXiv:1104.1787 [hep-th]];
 A.~Gadde and W.~Yan,
 arXiv:1104.2592 [hep-th];
 Y.~Imamura,
 JHEP {\bf 1109}, 133 (2011)
 [arXiv:1104.4482 [hep-th]].

\bibitem{Hama:2011ea}
 N.~Hama, K.~Hosomichi and S.~Lee,
 JHEP {\bf 1105}, 014 (2011)
 [arXiv:1102.4716 [hep-th]].

\bibitem{VolkovAbelian} 
 A.~Y.~Volkov,
 Phys.\ Lett.\ A {\bf 167}, 345 (1992)
 [hep-th/9307048];
 L.~D.~Faddeev and A.~Y.~Volkov,
 Phys.\ Lett.\ B {\bf 315}, 311 (1993)
 [hep-th/9307048];
 L.~D.~Faddeev,
 In {\it Varenna 1994, Quantum groups and their applications in physics}, 117-135
 [hep-th/9408041].

\bibitem{Bazhanov:2007mh}
 V.~V.~Bazhanov, V.~V.~Mangazeev and S.~M.~Sergeev,
 Nucl.\ Phys.\  B {\bf 784}, 234 (2007)
 [arXiv:hep-th/0703041].
 
 
 \bibitem{Bobenko}
  A.~Bobenko and B.~Springborn, Trans.\ Amer.\ Math.\ Soc.\ {\bf 356},
  659 (2004) [arXiv:math/0203250];
  B.~Springborn, arXiv:math/0312363.
 
 \bibitem{TQFT} 
 A.~Gadde, L.~Rastelli, S.~S.~Razamat and W.~Yan,
 Phys.\ Rev.\ Lett.\  {\bf 106}, 241602 (2011)
 [arXiv:1104.3850 [hep-th]];
 A.~Gadde, E.~Pomoni, L.~Rastelli and S.~S.~Razamat,
 JHEP {\bf 1003}, 032 (2010)
 [arXiv:0910.2225 [hep-th]].
  
\bibitem{3d3d} 
 Y.~Terashima and M.~Yamazaki,
 JHEP {\bf 1108}, 135 (2011)
 [arXiv:1103.5748 [hep-th]],  
 arXiv:1106.3066 [hep-th];
 T.~Dimofte and S.~Gukov,
 arXiv:1106.4550 [hep-th];
 T.~Dimofte, D.~Gaiotto and S.~Gukov,
 arXiv:1108.4389 [hep-th],
 arXiv:1112.5179 [hep-th]; 
 S.~Cecotti, C.~Cordova and C.~Vafa,
 arXiv:1110.2115 [hep-th];
 K.~Nagao, Y.~Terashima and M.~Yamazaki,
 arXiv:1112.3106 [math.GT].
\end{thebibliography}
\end{document}